\documentclass{article}                
\usepackage{graphicx,epsf,subfigure,pstricks,pst-node,psfrag,amsthm,amssymb,amsmath,url,inputenc}
\usepackage{float}
 
\begin{document}
\title{\textbf{\huge Mortality Rates of US Counties: Are they Reliable and Predictable?}}

\author{Robert L. Obenchain and S. Stanley Young}
\date{May 2023}
\maketitle

\begin{abstract}%
\noindent We examine US County-level \textit{observational data} on Lung Cancer mortality rates in 2012 and overall Circulatory Respiratory mortality rates in 2016 as well as their ``Top Ten'' potential causes from Federal or State sources. We find that these two mortality rates
for $2,812$ US Counties have remarkably little in common. Thus, for predictive modeling, we use a single \textit{compromise} measure of mortality that has several advantages. The vast majority of our new findings have simple implications that we illustrate graphically.
\end{abstract}



\maketitle

\section{Introduction} 

We use functions from CRAN-packages, R Core Team (2023), to apply cutting-edge statistical computing and graphical methods. The
``radon'' and ``pmdata'' data frames stored within the \textit{LocalControlStrategy} $R-$package, Obenchain (2022c), provide
detailed information for individual US Counties. When these frames of observational data are \textit{merged} using \textit{fips}-codes, data for a total of $2,812$ US Counties have no missing values for all variables listed in Table 2 on page 4. The ``radon''
data.frame was originally amassed by Kristic (2017), while the ``pmdata'' frame contains data from both Pye et al.
(2021) and CDC Wonder for the year 2016. The analyses described here update and extend those previously discussed in Obenchain,
Young and Krstic (2019) and Obenchain and Young (2023).

\section{US County Data Availability}
 
Mortality rates for individual US Counties collected during the years of 2012 through 2016 apparently provide the most recent rates
\textit{not impacted} by the COVID-19 pandemic. On the other hand, the (low) 2012 rates \textit{count only deaths due specifically to lung cancer}, while the higher 2016 rates of Circulatory and/or Respiratory mortality count deaths attributed to all relevant diseases
rather than cancers alone.

\begin{tabular}{@{}lcccccc@{}}
\multicolumn{7}{c}{\textbf{TABLE 1 -- Summary Statistics for lcanmort and CRmort}} \\
                  &                  & \textbf{First}    &                 &               &  \textbf{Third}   &  \\
\textbf{Variable} & \textbf{Minimum} & \textbf{Quartile} & \textbf{Median} & \textbf{Mean} & \textbf{Quartile} & \textbf{Maximum} \\
lcanmort (2012)   &   6.762          &  67.31            &  79.16          &  78.13        &  89.71            & 205.75 \\
CRmort (2016)     &   64.83          & 360.95            & 451.96          & 458.00        & 544.23            & 1564.08 \\
\end{tabular}
\vspace{0.5cm}

\noindent Note that 2012 cancer mortality rates generally exceed $10\%$ of their corresponding Circulatory-Respiratory mortality rate in 2016. Thus, 2012 \textit{cancer mortality rates} could be multiplied by a factor of $10$ simply to make them more comparable to 2016 rates of Circulatory-Respiratory mortality that already include cancers.

\section{Year-to-Year Variability in Mortality within US Counties}

Traditional methods for \textit{dimensionality reduction}, such as ``principal components'', could be used to define a
univariate \textit{compromise} measure of Circulatory-Respiratory mortality from the available 2012 and 2016 rates. Unfortunately,
the fundamental difficulty in doing this is that computed ``principal coordinates'' have (by definition) a mean-value of
\textit{ZERO}. Let $Zmin$ denote the \textit{absolute value} of the most-negative computed coordinate within a single component,
$Z$. Thus some numerical value strictly larger than $Zmin$ would need to be added to $Z$ to define a mortality ``rate'' consisting of strictly positive numerical values. There appears to be no clearly objective (or otherwise ``meaningful'' way) to do this.   

\begin{figure}[H]
\center{\includegraphics[width=5in]{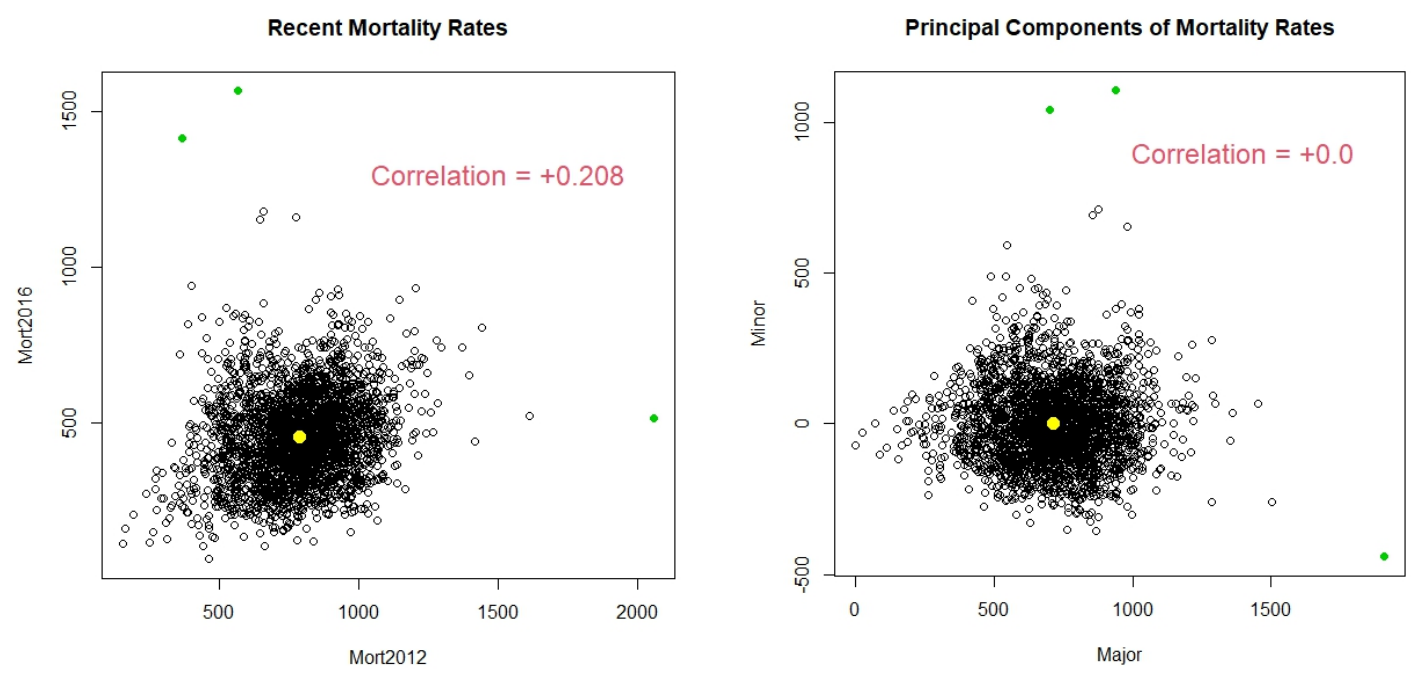}}
\caption{These ``scatter plots'' for $2,812$ US Counties within 46 States reveal considerable disagreement between 2012 and 2016
\textit{Mortality} rates per 100K residents. While the left-hand plot displays these rates, the right-hand plot shows their
Major (more variable) principal-coordinates on it's horizontal axis and their Minor principal-coordinates on it's vertical axis.}
\label{Fig01}  
\end{figure}

Key points are illustrated by the vertical (Minor) coordinates in the right-hand scatter of Figure 1 on page 2. We added $+714.6$ to the horizontal (Major) coordinates in the right-hand scatter simply because the minimum of the original computed Major coordinates was $-714.6$. Readers may find that the three relatively extreme points that are colored green in both scatters help them see similarities between these two scatters. Major (horizontal) coordinates in the right-hand scatter are much more strongly associated (correlation
$= 0.95$) with 2012 cancer mortality rates than with 2016 CR mortality rates (correlation $= 0.49$).

\subsection{A Simple ``Compromise'' Mortality Rate}

The ``compromise'' measure of mortality, $Cmort$, that we will use in our analyses is simply the \textit{average} of [1] the
relatively high 2016 $CRmort$ rates and [2] the relatively low $lcanmort$ rates of $2012$ multiplied by $10$. This definition is
simple, (hopefully) intuitive, and yields $2,812$ strictly positive mortality rates. The pair of plots displayed in Figure 2
below show how these new $Cmort$ rates compare with the rescaled $2012$ and raw $2016$ mortality rates.

\begin{figure}[H]
\center{\includegraphics[width=5in]{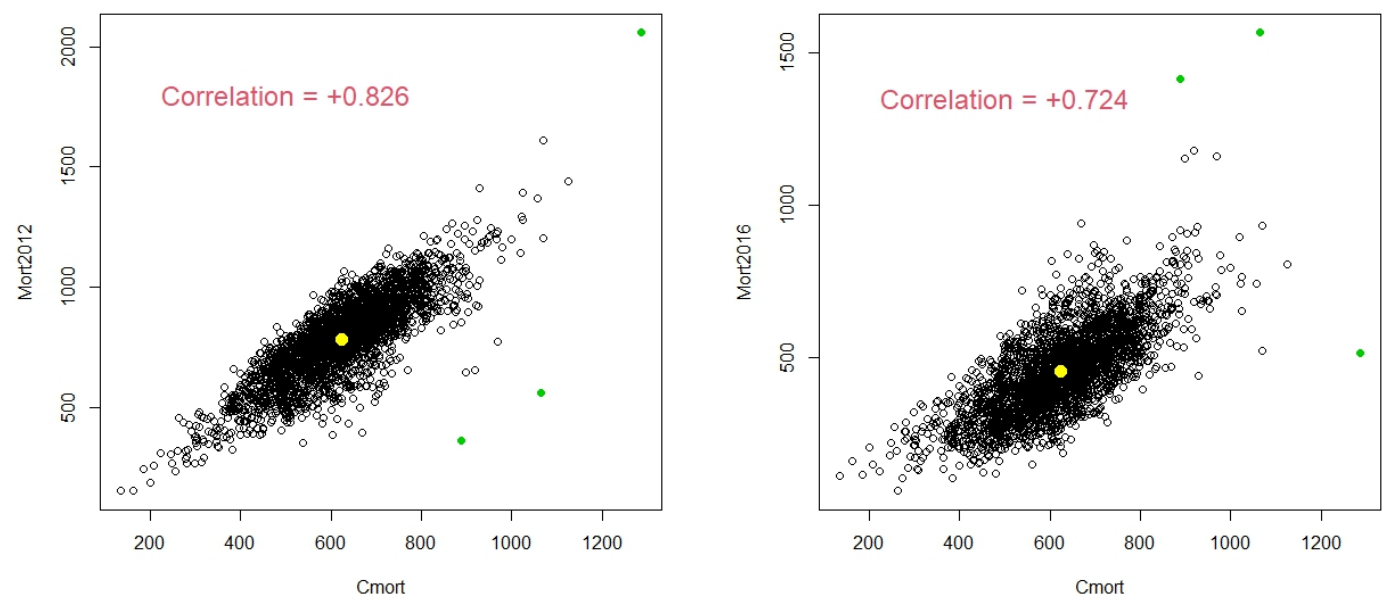}}
\caption{These scatter plots for $2,812$ US Counties show how our ``compromise'' rate, $Cmort$, of mortality per 100K
residents compare with $2012$ and $2016$ rates. The same three US Counties with relatively extreme rates in $2012$
or $2016$ that were colored green in Figure 1 are also colored green here. Note that the $Cmort$ rate is only slightly more highly
associated with the $2012$ lung cancer rate than with $2016$ ``Circulatory and/or Respiratory'' mortality.}
\label{Fig18}  
\end{figure}

Our new analyses will focus on alternative models for prediction of these compromise $Cmort$ rates. Our ``Top Ten'' explanatory
$X-$variables are described in the $10$ final rows of Table 2. In Obenchain, Young and Krstic (2019), we used a logarithmic
transformation of $Radon$ that we avoid here. Thus, there will be no need to ``Winsorize'' the $Radon = 0.0$ levels reported
for $10$ US Counties.

\vspace{0.5cm}
\begin{tabular}{@{}lll@{}}
\multicolumn{3}{c}{\textbf{TABLE 2 -- Variable Information}} \\
\textbf{Name} & \textbf{Description} & \textbf{Range} \\
fips & Federal Information Processing Std code ($4$ or $5$ digits) & 1001-56043 \\
State & Two Character State ID Code & ``AL'' - ``WY'' \\
County & County Name (Character String) &  \\
lcanmort & 2012 Lung Cancer Mortality / 100K Residents & $6.762-205.75$ \\
CRmort & 2016 Circulatory-Respiratory Mortality / 100K Residents & $64.8-1,564.1$ \\
Cmort & ``Compromise'' Mortality / 100K Residents & $135.2 - 1,286.3$ \\
Smoking & Percentage of Residents who Currently Smoke & $7.3\% - 40.9\%$ \\
Elderly & Percentage of Residents Over 65 & $3.0\% - 34.7\%$ \\ 
Radon & Average Indoor Radon Level in pica-Curies per Liter & $0.0 - 99.7$ \\   
NO2 & Nitrogen Dioxide Percentage &  $0.15\% - 19.6\%$ \\     
Ozone & O3 Percentage & $19.5\% - 41.1\%$ \\     
Sulfates & SO4 Percentage in Particulate Matter & $0.39\% - 1.6\%$ \\  
PremDeath & Premature Death Rate & $2,853 - 36,469$ \\
ChildPov & Children Living in Poverty & $2.9\% - 66.3\%$ \\  
Avoc & Anthropocentric Volatile Organic Compounds & $0.228 - 2.891 \mu{g}/m^3$ \\ 
Bvoc	 & Biogenic Volatile Organic Compounds & $0.261 - 3.309 \mu{g}/m^3$ \\
\end{tabular}

\subsection{Predicting Mortality using a Single Potentially Causal Variable} 

\begin{figure}[H]
\center{\includegraphics[width=3in]{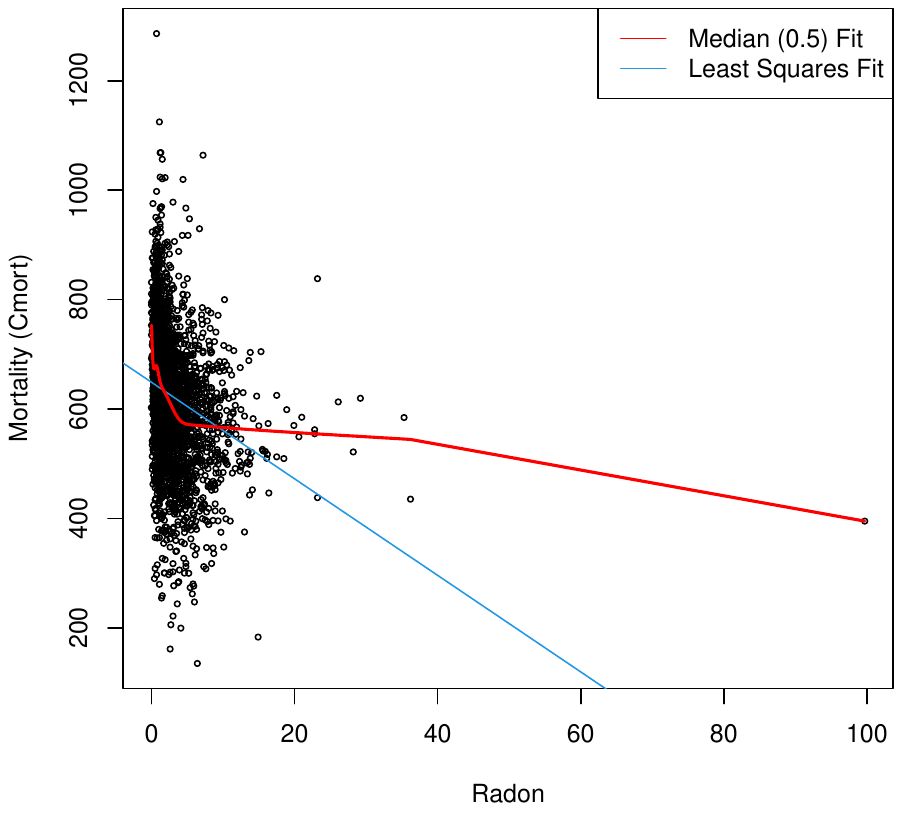}}
\caption{This scatter for $2,812$ US Counties within 46 States again reveals high variability within our ``compromise'' \textit{Cmort} rates per 100K residents. While $96$ Counties experience average radon levels greater than 10 $pCi/L$, only 13 of these levels exceed
20 $pCi/L$. The red curve shows the \textit{median} regression quantile fit using the \textit{quantreg} $R-$package, Koenker
(2005-2022), with 9 degrees-of-freedom. Note that considerable non-linearity occurs for radon levels less than 35 $pCi/L$. Although
the data point for Teller County, CO, [$Cmort = 299.6$ per 100K residents at $radon = 99.7$ $pCi/L$] clearly has ``high leverage'',
this ``outlier'' is none-the-less essentially ignored by the linear model fit (blue line). In fact, no model for predicting Cmort
from radon levels that \textit{assumes error-terms are uncorrelated and homoscedastic} would seem fully appropriate here.}
\label{Fig02}  
\end{figure}

A good place to start our discussion of alternative ways to predict our $y-$outcome, $Cmort$, is provided by Figure~\ref{Fig02}.
We start with an example that uses a single potentially causal $x-$variable because both the input data and the resulting
predictions can then be displayed in two dimensions. Horizontal coordinates in Figure~\ref{Fig02} below are indoor $Radon$ levels
in pica-Curies per Liter (pCi/L).

Our primary ``Take Away'' from Figure~\ref{Fig02} is simply that modern statistical methods for fitting predictive models that are both
non-linear and highly flexible tend to provide much more realistic and practically useful predictions than rigid traditional methods
that assume linearity.

\subsection{Multiple Linear Regression Models using our Top Ten Predictors} 

\begin{figure} [H]
\center{\includegraphics[width=3in]{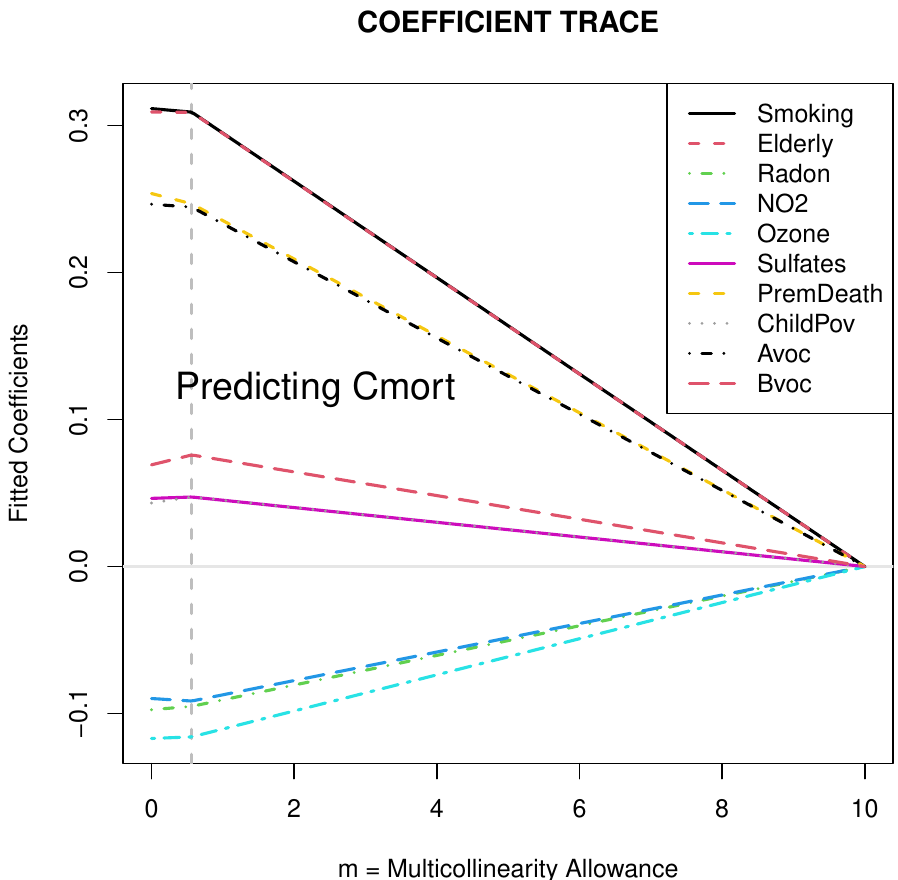}}
\caption{MSE-risk optimal relative-magnitudes for $\beta-$coefficient estimates are determined, Obenchain (2022a), by Shrinkage to
$m=0.6$ on the horizontal axis. In other words, overall ill-conditioning (redundancy) among these ten $X-$variables reduces
their linear modeling ``rank'' to essentially $9.4$. Note that four pairs of almost identical fitted coefficients emerge. Thus
$Smoking$ and $Elderly$ have the most positive \textit{linear effects} on $Cmort$ outcomes. $PremDeath$ and $Avoc$ have the next
most positive effect. Less positive \textit{linear effects} are attributed to $Bvoc$ as well as to the $Sulfates$ and $ChildPov$
pair. The most negative \textit{linear effect} on $Cmort$ outcomes from the $Ozone$ predictor, while the negative effects of the
$Radon$ and $Nitrogen Dioxide$ pair are somewhat less strong.}  
\label{Fig03} 
\end{figure}

Although we will ultimately focus on ``Black Box'' methods that are highly-flexible, let us start by fitting a \textit{multiple linear
regression} simply to assess the overall ``extent of ill-conditioning'' (confounding) among our ten potential $X-$predictor variables.
The Generalized Ridge Regression \textit{Trace} display in Figure~\ref{Fig03} shows regression $\beta-$coefficient estimates from the
eff.ridge() function within the \textit{RXshrink} R-package of  Obenchain (2022b). This family of linear models that are fit
using \textit{maximum-likelihood under Normal distribution-theory} attempt to predict $Cmort$ rates using our ``Top Ten'' potentially causal-variables $X-$variables. This family of estimators is indexed by the scalar parameter $m$ that varies continuously:
$0 \leq m \leq 10$, where $m = 0$ is the traditional least-squares solution and all estimates are shrunken to zero at $m = 10$.  

Due to missing values in the $PremDeath$ predictor variable for $3$ US Counties, these models and all remaining models for predicting $Cmort$ rates will be based upon data from only $2,812$ US Counties. The ``Black Box'' models of Section $4$ suggest that $PremDeath$
is easily the single most important of our ``Top Ten'' potentially causal determinants of $Cmort$ rates.

Our \textit{Random Forest} analyses described next will yield ``Black Box'' models that are \textit{distinctly Non-Linear}, make
minimal realistic assumptions, and provide much improved predictions of Mortality. In other words, we will see that several of the relative magnitudes of the simplistic $\beta-$coefficient estimates in Figure 3 are rather misleading. For example, all
$\beta-$coefficient estimates for $Bvoc$ (Red long-dash) are positive but smaller than those for $Avoc$ (Black dot-dash) coefficients
in Figure 3, but $Bvoc$ will prove to be more predictive of $Cmort$ than is $Avoc$ with \textit{non-linear} (more realistic) models.

Our ``Take Away'' from Figure~\ref{Fig03} is simply that ill-conditioning (inter-correlations among) our ``Top Ten'' potentially
causal variables is apparently \textit{not a serious problem}.

\section{Our Non-parametric Supervised Learning ``Black Box'' Model}

We generated a random forest of $500$ tree models, Breiman (2001,2002), for prediction of $Cmort$ from our ``Top Ten''
potentially causal X-variables using default settings in the randomForest $R-$package of Liaw and Wiener (2002-2022). Using the
corresponding Partial Dependence Plots, Friedman (2001), that are both generated and ranked on ``importance'' by this software,
we studied the ten \textit{marginal relationships} that result from averaging over the other nine potential predictors. These
marginal relationships ignore potential interaction effects and can be linear, monotonic or more complex.

In addition to a Partial Dependance Plot (PDP), Figures $5$ to $14$ also display a companion \textit{Individual Conditional Expectation} (ICE) plot, Goldstein et al. (2015). The ``rugs'' (nine vertical tick-marks extending above the bottom axis) on each PDP plot mark boundaries between observed deciles for each potential predictor of $Cmort$. On the corresponding ICE-plot, note that these deciles are \textit{uniformly} spaced. Thus, individual ICE-plots transform the horizontal range displayed by its companion PDP.

Table 3 summarizes PDP characteristics of our $10$ pairs of PDP and ICE plots in Figures $5$ to $14$.

\vspace{0.5cm}
\begin{tabular}{@{}ccc@{}}
\multicolumn{3}{c}{\textbf{TABLE 3 -- Importance Statistics for the ``Top Ten'' predictors of Cmort Rates}} \\
\textbf{Variable}  &  \textbf{\%IncMSE}   &  \textbf{IncNodePurity} \\
PremDeath & 53.73973  & 9752125 \\
Elderly   & 48.96057  & 3557815 \\
Smoking   & 43.43565  & 5740271 \\
Bvoc      & 39.45565  & 5798775 \\
Ozone     & 34.36222  & 2487499 \\
ChildPov  & 32.29134  & 4976096 \\
Avoc      & 28.95409  & 3349377 \\
Sulfates  & 27.97395  & 2739202 \\
NO2       & 25.06410  & 1366603 \\
Radon     & 19.69432  & 1408827 \\
\end{tabular}

\begin{figure}[H]
\center{\includegraphics[width=5in]{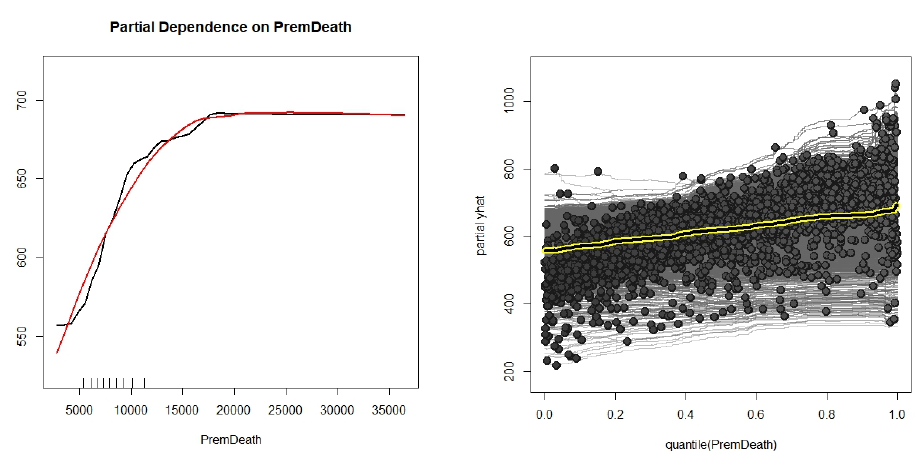}}
\caption{\label{Fig04} The most important predictor of $Cmort$ is the $Premature Death$ rate among county residents:
$\%$IncMSE$=53.7$. The relationship between the $Premature Death$ rate and $Cmort$ is essentially Monotone Increasing.
Only one US County has a $Premature Death$ rate $> 20,000$. Minimum rates of $5,000$ or less appear to be highly desirable!}
\end{figure}

\begin{figure}[H]
\center{\includegraphics[width=5in]{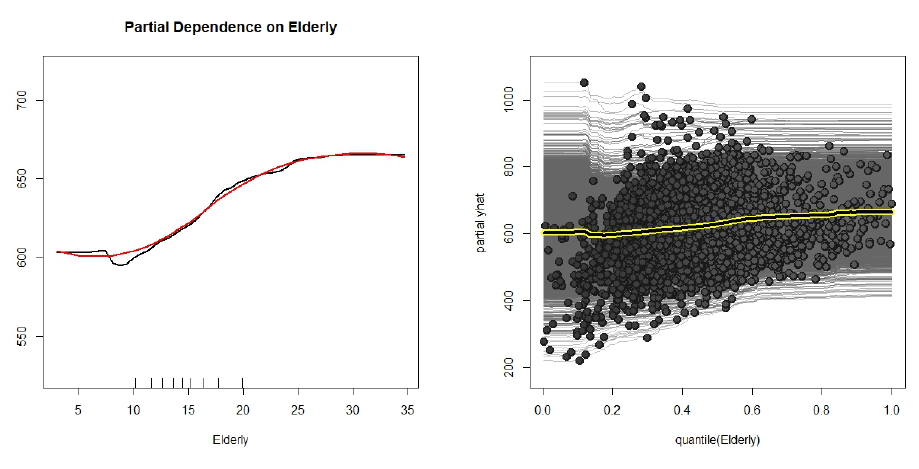}}
\caption{\label{Fig05} The second most important predictor of $Cmort$ is each County's percentage of $Elderly$ residents:
$\%$IncMSE$=49.0$. The relationship between the $Elderly$ percentage and $Cmort$ appears to be Monotone Increasing when this
percentage exceeds $10\%$. }
\end{figure}

\begin{figure}[H]
\center{\includegraphics[width=5in]{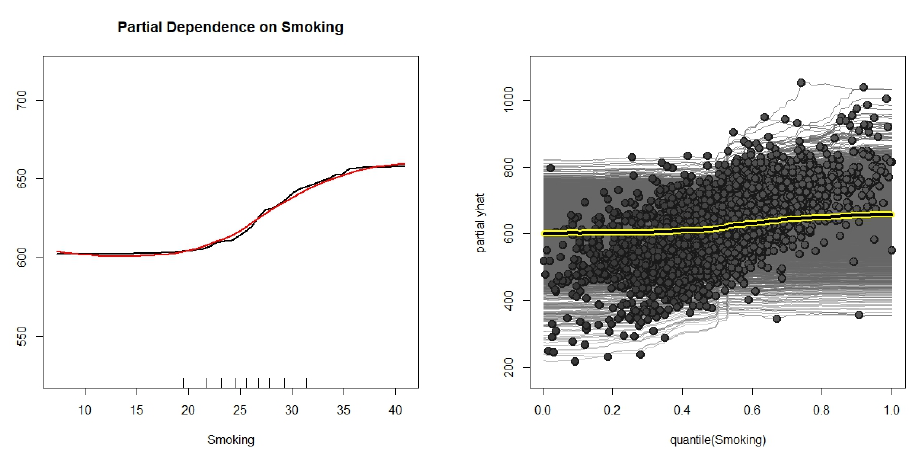}}
\caption{\label{Fig06} The third most important predictor of $Cmort$ is each County's $Smoking$ percentage : $\%$IncMSE$=43.4$.
While the relationship between $Smoking$ and $Cmort$ appears to be Monotone Increasing on its ICE plot, the trend is initially
rather flat for $Smoking \leq 18\%$, where only $10\%$ of US Counties provide data.}  
\end{figure}

\begin{figure}[H]
\center{\includegraphics[width=5in]{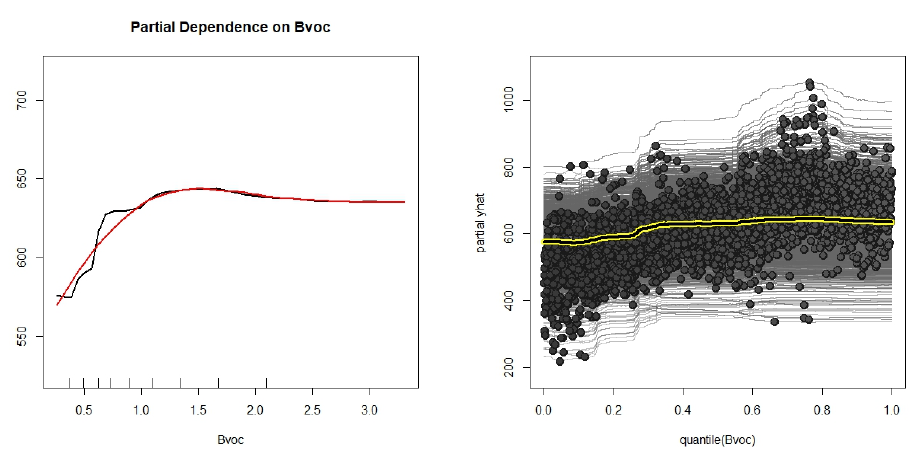}}
\caption{\label{Fig07} The fourth most important predictor of $Cmort$ is each County's level of ``Biogenic (Natural)
volatile organic compounds'' : $\%$IncMSE$=39.5$. Note that $Cmort$ rates sharply increase until $Bvoc$ levels reach
roughly 1.5 $\mu{g}/m^3$ ...then decrease somewhat for higher $Bvoc$ levels where only $20\%$ of US Counties provide data.} 
\end{figure}

\begin{figure}[H]
\center{\includegraphics[width=5in]{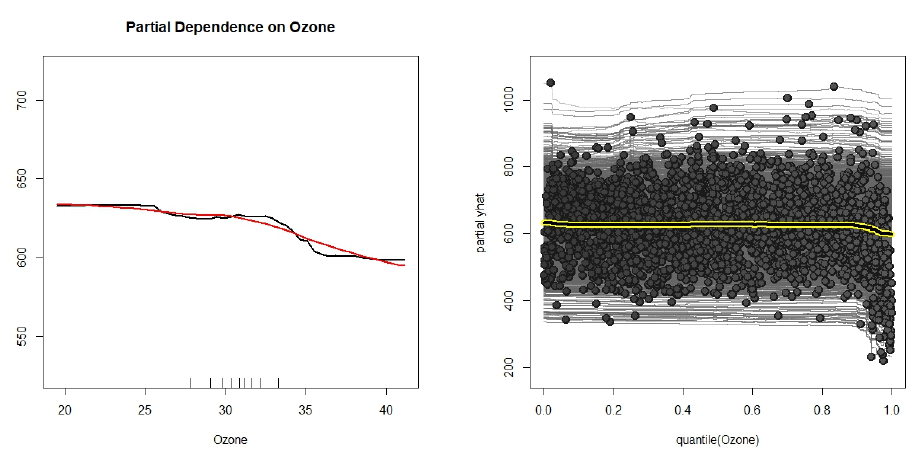}}
\caption{\label{Fig08} The fifth most important predictor of $Cmort$ is each County's $Ozone$ level, $\%$IncMSE$=34.4$.
Surprisingly, $Cmort$ decreases monotonically as $Ozone$ levels increase!} 
\end{figure}

\begin{figure}[H]
\center{\includegraphics[width=5in]{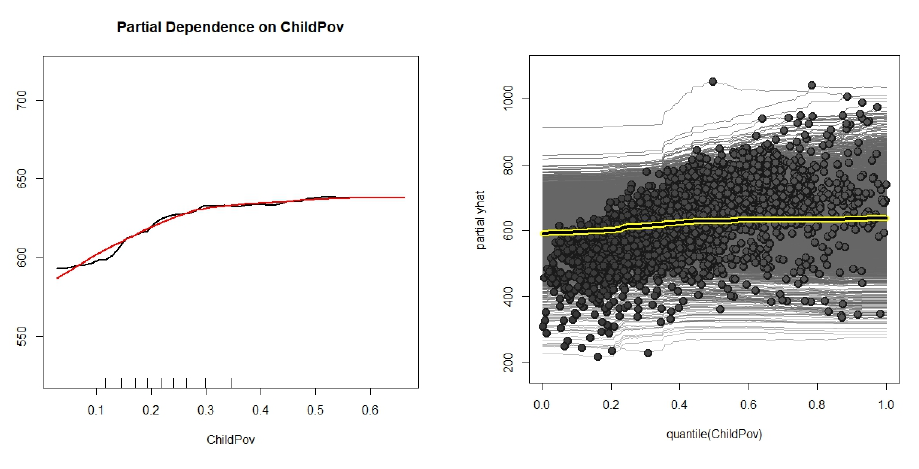}}
\caption{\label{Fig09} A sixth important predictor of $Cmort$ is each County's percentage of ``Children Living in Poverty'' :
$\%$IncMSE$=32.3$. Furthermore, $Cmort$ rates appear to increase monotonically with this percentage.}
\end{figure}

\begin{figure}[H]
\center{\includegraphics[width=5in]{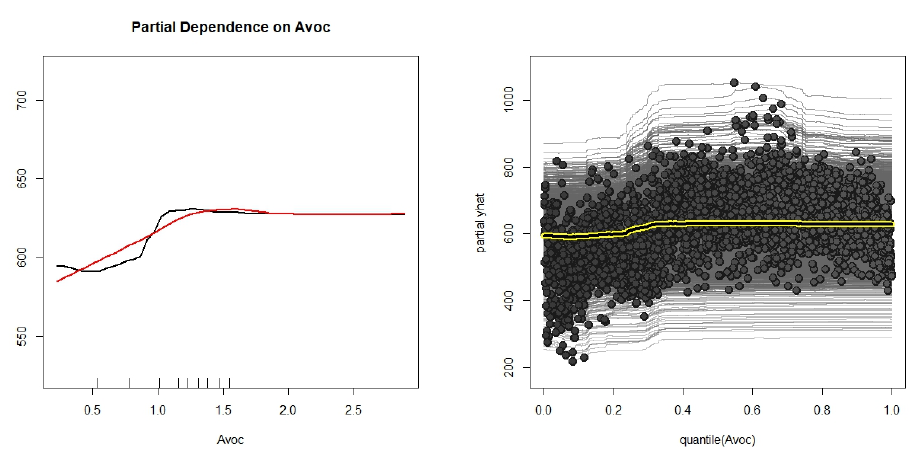}}
\caption{\label{Fig10} Another key predictor of $Cmort$ is each County's $Avoc$ level of ``Anthropocentric (Man-Made)
volatile organic compounds'', $\%$IncMSE$=29.0$. $Cmort$ rates level-off when $Avoc$ levels exceed $1.5 \mu{g}/m^3$.
This pattern is much like that of $Bvoc$ effects.}
\end{figure}

\begin{figure}[H]
\center{\includegraphics[width=5in]{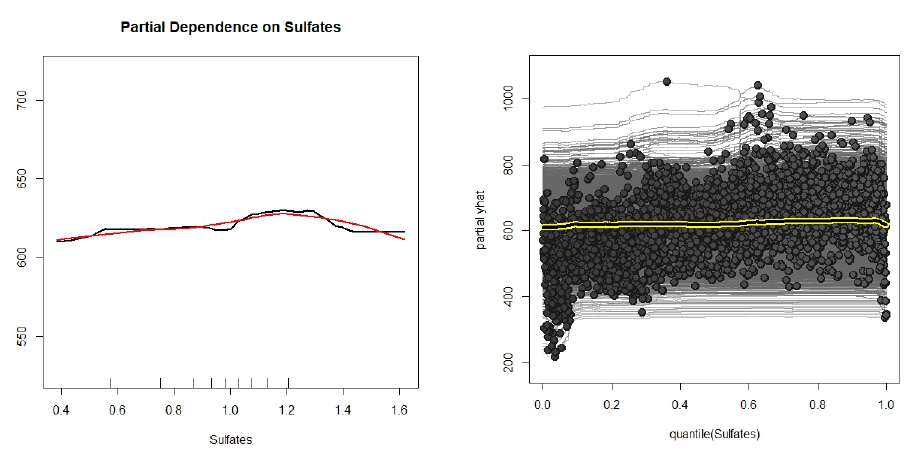}}
\caption{\label{Fig11} Another important predictor of $Cmort$ is each County's $Sulfate$ level, $\%$IncMSE$=28.0$. Note that
$Cmort$ rates do not monotonically decrease or increase as the percentage of air-borne $Sulfate$ increases. Instead, $Cmort$
rates appear to peak somewhere within $1.1\% < SO4 < 1.3\%$. Only about $10\%$ of US Counties have $SO4 > 1.2\%$. Finally,
the correlation ($+0.7185$) between $Avoc$ levels in Figure~\ref{Fig10} and $Sulfate$ levels is larger than that between any other
pair of potentially causal predictors of mortality.}
\end{figure}

\begin{figure}[H]
\center{\includegraphics[width=5in]{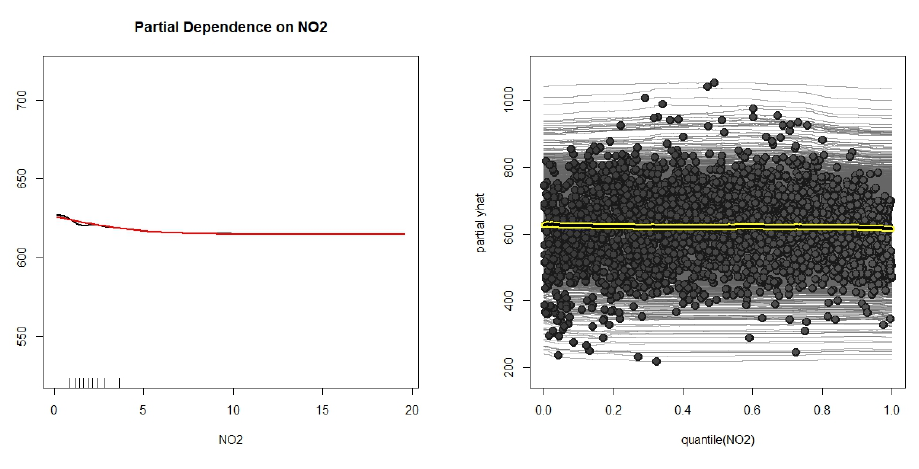}}
\caption{\label{Fig12} Another important predictor of $Cmort$ is each County's Nitrogen Dioxide ($NO2$) level, $\%$IncMSE$=25.1$.
Note here that $Cmort$ rates are nearly constant at all but the very lowest observed NO2 levels. Specifically, the $90\%$ of US Counties with $NO2$ levels below roughly $4\%$ tend to have slightly higher $Cmort$ rates.}
\end{figure}

\begin{figure}[H]
\center{\includegraphics[width=5in]{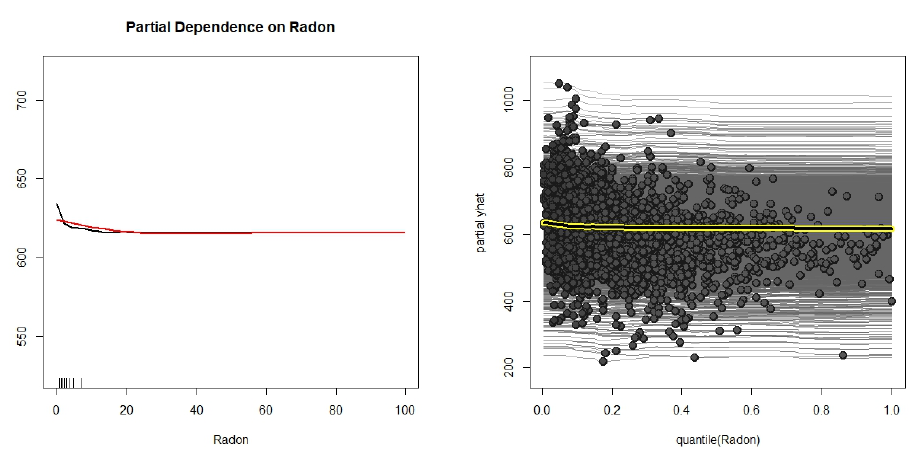}}
\caption{\label{Fig13} The final ``Top Ten'' predictor of $Cmort$ is each County's indoor Radon level : $\%$IncMSE$=19.7$.
Clearly, $Cmort$ rates drop when indoor $Radon$ levels exceed $8$ pica-Curies per Liter. Unfortunately, only residences in
about $10\%$ of US Counties have indoor Radon level this high. Note that the companion ICE-plot confirms that this
relationship is monotone decreasing.} 
\end{figure}

In summary, our PDP and ICE plots display ten potentially causal predictors of $Cmort$ rates; all of these relationships are
distinctly non-linear. Two of these Ten predictors concern aging ($Elderly$ and $PremDeath$), two more are socioeconomic factors
($Smoking$ and $ChildPov$), while the other six concern chemicals in the air.

\section{Four Main Factors associated with Mortality}   

Our pairs of PDP and ICE plots in Figures \ref{Fig04} to~\ref{Fig13} have encouraged us to group our ``Top Ten'' predictors into
just four main ``Factors'' that impact the health of the US population: [1] Longevity, [2] Socio-Economic Issues, [3] Regulated
Chemicals in indoor or outdoor Air and [4] Volatile Organic Compounds within Air Pollution [detectable via satellite images].

\subsection{Many Effects of Longevity are Obvious}

It certainly is not surprising that the most important predictor of $Mortality$ turned out to be each County's percentage of $Elderly$ residents (age 65 or over). All we have learned here is that this relationship is \textit{not purely linear}; i.e. genetic factors may help some people age more ``gently'' than others. Perhaps Metformin or future medicines will actually help increase longevity; see
Barzilai et al. (2016).

\subsection{Socio-Economic Effects}

The Second, Third and Fifth most important predictors of $Mortality$ are [2] the Premature-Death Rate, [3] (Adult) Smoking Percentages
and [5] Children Living in Poverty. While many Socio-Economic problems have rather clear effects, what is \textit{not clear} is how to most effectively change them among citizens of a constitutional republic. 

\subsection{Effects of Regulated Chemicals}
The predictors ranked Seventh through Tenth here are: [7] Ozone ($O3$), [8] Sulfates ($SO4$), [9] Nitrogen Dioxide ($NO2$), and [10]
Radon ($Rn$). Both the general US population and EPA regulations essentially \textit{assume} that these chemical elements in the air we breathe have highly detrimental effects. However, all but Sulfates have PDP and ICE profiles showing that $Mortality$ tends to \textit{decrease monotonically as their presence increases}! For Sulfates, the PDP profile appears to peak somewhere near $SO4 = 1.2$;
see Figure 12.

The $Radon$ profile depicted in Figure 13 strongly re-enforces our published findings, Obenchain, Young and Krstic (2019). We fear
that many humans harbor an almost irrational fear of even rather low doses of \textit{ionizing radiation}. Thus, we now focus on $Rn$
in indoor air even though both of the ``more important'' predictors in outdoor air, $O3$ and $NO2$, have the essentially same profiles
as $Rn$.  

Figure 15 depicts clear differences between three types of potential relationships between the intensity of \textit{ionizing radiation}
and its effects on Mortality rates: [1] a ``Linear No Threshold'' (LNT) model, [2] a three-part model (first``no effect''
then a ``sharp threshold effect'' followed by a ``linearly increasing trend''), and [3] a \textit{fully realistic} ``radiation hormesis'' model under which undesirable radiation levels are either too low or else much too high. Our \textit{Random Forest of Tree Models} generated the profile depicted in Figure 14 on page $12$ that implies hormesis.

\begin{figure}[H]
\center{\includegraphics[width=6in]{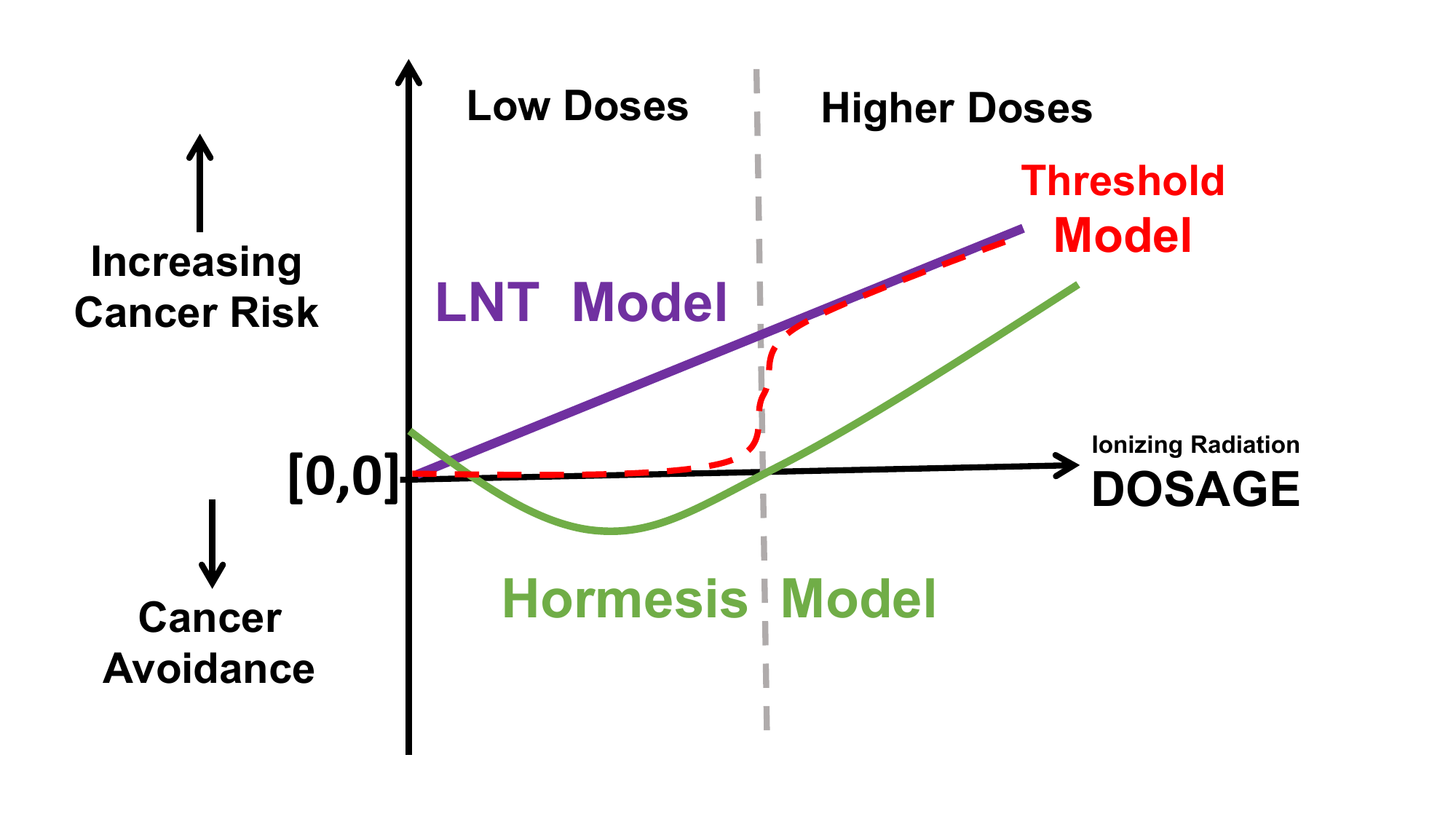}}
\caption{\label{Fig14} This figure depicts important differences between [1] the widely accepted \textit{Linear No Threshold} (LNT) model (Purple) for the effects of ionizing radiation on mortality, [2] a model with a distinct threshold (Red), and [3] an ``Ionizing Radiation Hormesis'' model (Green) that is \textit{fully consistent} with the analyses presented here.}
\end{figure}

Many published papers support $Rn$ hormesis. Cohen (1989,1995,1997,2008) maintained that the most desirable indoor $Radon$ levels are between 7 and 10 $pCi/L$; also, see Parsons (2002) and Calabrese et al. (2007). More recent work, Castillo et al. (2015) and (2017),
notes that bacteria deprived of background levels of ionizing radiation suffer a \textit{stress response}; this more recent work
signals actual causation, rather than mere association.

\begin{figure}[H]
\center{\includegraphics[width=5in]{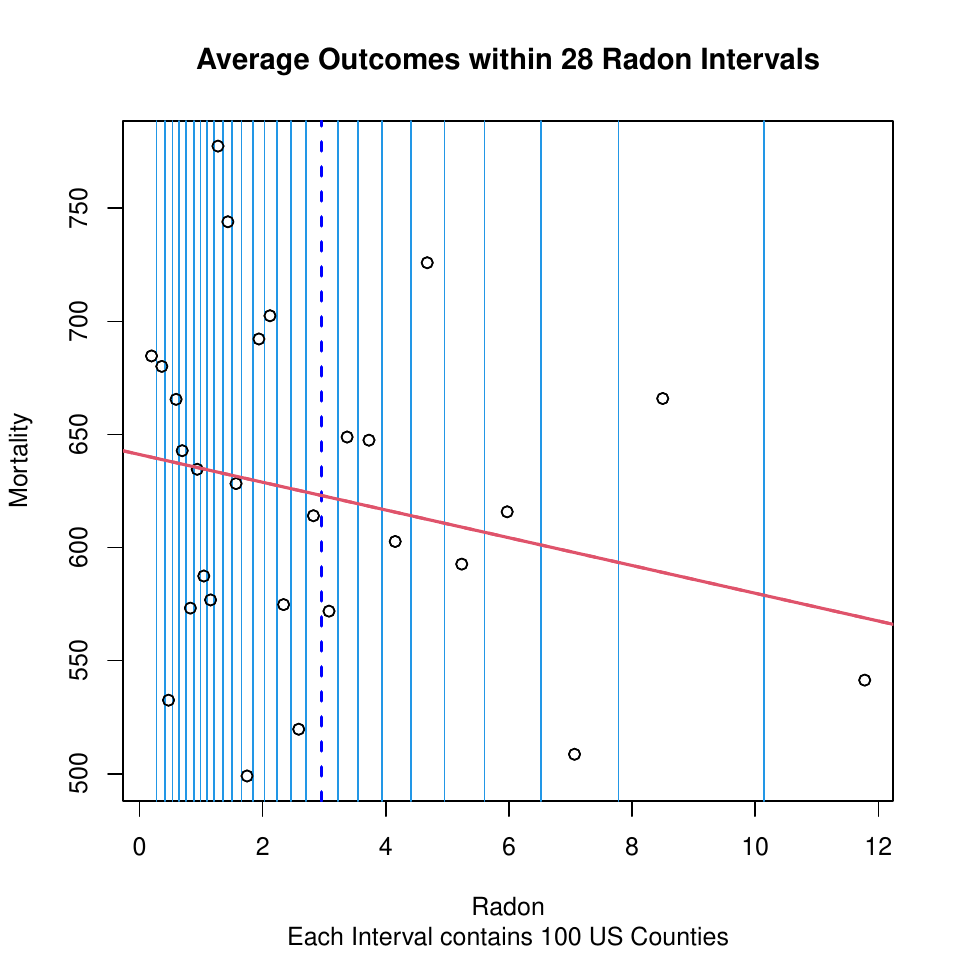}}
\caption{\label{Fig15} This figure shows how average $Mortality$ counts within $28$ ordered subgroups (each containing $100$ US Counties) tend to \textit{decrease} as their Radon levels increase! The darker-blue dashed line at Radon $= 2.95$ $pCi/L$ corresponds to the mean exposure for the given $2,800$ lowest exposures. The 12 remaining Counties with highest Radon levels have an average Mortality rate of $705.6$. Unfortunately, the US EPA currently recommends ``Radon Mitigation'' for exposures \textit{greater than 4.0 pica-Curies per Liter}, and many US States enforce this expensive (and generally counter-productive) requirement.}
\end{figure}

\subsection{Effects of Secondary Organic Aerosols}

Pye et al. (2021) stated: ``Underlying medical conditions like heart disease as well as CR [Circulatory/Respiratory] disease
mortality are also higher in the Southeast than the rest of the U.S. as a result of multiple socioeconomic and behavioral factors.''

We ask: What is it that US Counties with an abundance of Bvoc consistently lack? By comparing Figures 15 and 16, we have come to the
following conclusion: \textit{Most lack sufficient indoor Radon levels to benefit from ionizing radiation hormesis!}

\begin{figure}[H]
\center{\includegraphics[width=6in]{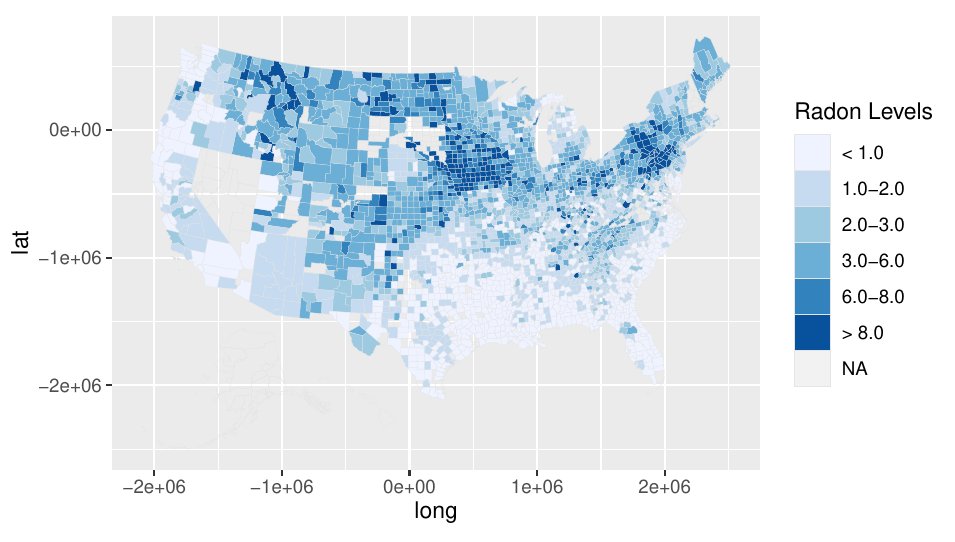}}
\caption{\label{Fig16} This US Map shows variation in indoor \textit{radon levels} for $2,812$ counties ...using $6$ shades of blue. Since no counties within the States of Nevada and New Hampshire are included in our \textit{radon} data.frame, these two entire States receive the ``see-through'' background color.}
\end{figure}

\begin{figure}[H]
\center{\includegraphics[width=6in]{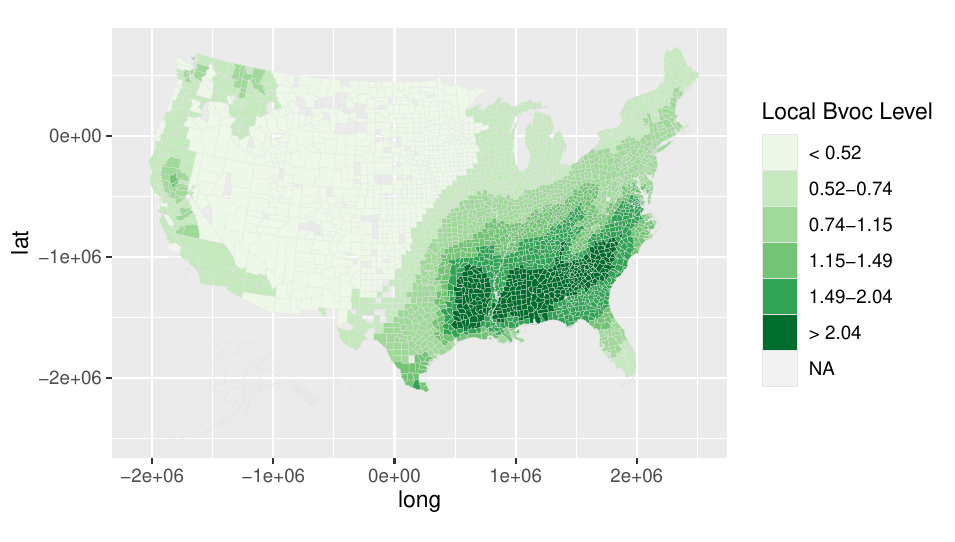}}
\caption{\label{Fig17} This US Map shows variation in \textit{Bvoc} levels for $2,881$ counties ...using $6$ shades of green. Since most
southeastern US states (other than southern Florida) not only have particularly high \textit{Bvoc} levels but also quite \textit{low radon levels} in Figure 15, their Mortality rates fail to benefit from Radon hormesis.}
\end{figure}

\section{Summary}

We have illustrated \textit{Statistical Learning} using a Random Forest of Tree models for analysis of cross-sectional \textit{observational data} from $2,812$ US Counties. This approach helps readers literally ``see'' potentially causal
marginal relationships between our \textit{Top Ten} explanatory variables and Mortality outcomes. Our analyses
pay special attention to indoor Radon levels because, contrary to the long-standing EPA requirement that Radon levels
must be below 4.0 pica-Curies per Liter, we confirm that Mortality rates tend to actually be \textit{lower} for US Counties
with average indoor Radon levels \textit{above} this threshold than below it. In other words, \textit{ionizing radiation}
tends to be beneficial (\textit{hormetic}) rather than detrimental to human health. On the other hand, our analyses also
confirm that Secondary Organic Aerosols in Air Pollution appear to consistently increase Mortality rates.

To reduce $Mortality$ rates, desirable indoor $Radon$ levels appear to be not only above roughly 4 $pCi/L$ but also
below roughly 30 $pCi/L$. In other words, undesirable indoor Randon levels can be either too low or much too high.

Our PDP and ICE plots for the ``Top Ten'' predictors of Mortality yield Random Forest ``Black Box'' models that provide much more realistic information than can be gained simply by examining $\beta-$coefficient estimates from multiple linear regression, as in
Figure 3. True effects among the many potentially relevant variables considered here are definitely more complex than simple
linear models with homoscedastic measurement errors! Furthermore, these new \textit{Supervised} learning results appear to be
compatible with those from the Non-parametric \textit{Unsupervised} approach of Obenchain and Young (2023).

It certainly appears worthwhile to collect and report mortality rates for individual US States and Counties, and it does
``sound good'' that many of these rates have apparently been decreasing recently. On the other hand, we have seen that these sources
of \textit{Observational Data} tend to be much too highly variable (both across locations and year-to-year) to be truly valuable in
making accurate comparisons and forecasts.

\subsection*{Conflict of Interest}

As independent and self-funded researchers, the authors declare that no competing interests exist.

\end{document}